\documentclass[12pt]{article}
\usepackage{amssymb}
\usepackage{amsmath}
\usepackage{epsfig}
\newcommand{\be}{\begin{equation}}
\newcommand{\ee}{\end{equation}}
\newcommand{\bea}{\begin{eqnarray}}
\newcommand{\eea}{\end{eqnarray}}

\usepackage{epsfig}
\usepackage{array}
\usepackage{amsmath}
\usepackage{amssymb}
\usepackage{graphicx}
\newcommand{\Chi}{\mathrm X}

\title {HIDDEN VARIABLES IN ANGULAR CORRELATIONS OF THE PARTICLES EMITTED IN FISSION  }
\author{ F. F. Karpeshin \\ D. I. Mendeleev All-Russian Research
Institute for Metrology \\ 190005 Saint-Petersburg, Russia }
\date{}

\begin{document}
\maketitle

\begin{center}
 \vspace{0.1cm}
{\it E-mail: fkarpeshin@gmail.com}
\\
\end{center}

\abstract{ A comparative analysis of experiments on studying the $ (n, f) $, on  one hand, and $ (n, n) $, on the other hand,  angular correlations in fission is carried out, based on the model proposed by muonic conversion in fragments of prompt fission of $^{238}$U with negative muons. Their fundamental difference is shown in the sense of the information that can be inferred from them. Historically, the situation resembles the EPR paradox. An experimental check of the empirical relation between the alignment and polarization parameters
$ A_ {n, J} = 2 A_{n, f}$ is proposed. }
\normalsize

\section {Introduction}

Probably, the most revolutionary moment in the development of
quantum mechanics (QM) was the Heisenberg uncertainty relation. It showed that far not each of the parameters one uses in the everyday life can be also measured in microcosm with an arbitrary precision. This seemed paradoxical from the point
of view of everyday consciousness. The famous
Einstein's expression ``God does not play dice'' and the EPR paradox, the
interest in which continues to this day, date back to this time.
To test and understand the principles of QM, the
theory of ``hidden parameters'' was put forward. It runs that
there may be additional parameters, for example, spin projection.
If we knew them before measurement, we could predict the result
unambiguously. Indeed, classical spin naturally has three projections. In quantum mechanics, there is only one left --- that on the quantization axis. The other two remain quite objective parameters, but pass into the area of those hidden from observation. If someone wants, nevertheless, to deal with them explicitly, then he will receive the wrong answer.

  The HV-theory is rejected by the community. However, examples can be noted of how variations on the theme of HV suddenly interplay, for example, in modern simulations of the angular distributions of gamma quanta or neutrons emitted from fission fragments. Consider the angular distribution of the neutrons from fission with respect to the direction of the light fragment. It is well known that the spins of fragments are mainly  aligned in a plane perpendicular to the fission axis. In order to take this circumstance into account,  one may consider the spin of each fragment to have a definite direction in the plane perpendicular to the fission axis, and then average over the directions in the plane.
Nevertheless, such a way that might seem evident at first glance, turns out to be certainly erroneous.
In such an approach, the supposed direction of the fragment's spin appears as a HV, additional to the projection on the quantization axis $z$ which is in the fission direction. Contrary, in a consecutive QM approach, the state of the fragment is characterized by two quantum numbers: the spin and its projection onto the quantization axis $z$. Then the alignment of the fragments merely means that the projection of their spins onto this axis is close to zero. And in the general case of incomplete alignment, it is necessary to use the density matrix.

       Therebefore, the angular correlations provide a useful guide to
experimental studying the interrelation of hidden variables with
QM. An explicit consideration of the additional
spin projection, which is treated as a non-observable one in a
consistent QM approach, leads to an obvious
contradiction with QM, which can be observed
experimentally. Let us consider this interplay in more
detail. We use angular correlations of the conversion muons,
emitted from the fission fragments in prompt fission of actinide
nuclei in muonic atoms of $^{233}$U, occurring as a result of
radiationless excitation in a muonic transition. This process was
discovered in JINR \cite{polk}. We remind the process in the next
section. Formulas necessary for analysis of the angular
correlations in the c. m. of the fragments are derived in section \ref{corl}. The account of the translation of the muon on the fragment is performed in section \ref{contradic}. The muons
emitted as a result of shake-off are similar to the scission
neutrons emitted in fission. Their contribution is taken into account when  comparing with experiment.
 In the concluding section, we sum up the
results obtained, draw conclusions, outline prospects for future
investigation.

\section{Physical premises}
\label{polk}

This unique process gives a direct information on the fission
dynamics. Beams of negative muons are slowed down in matter, then
the process of the muon capture into high orbits of the muonic
atoms starts. The muons are captured into the orbits with the main
quantum number $n\approx 14$. After this, they cascade down by means of
the radiative and Auger transitions. When the muons jump between
the lowest inner orbits, there is a chance that the transition
energy is transferred to the nucleus, which undergoes fission.
This kind of fission is called prompt fission, contrary to the
delayed fission, induced by the nuclear capture of the muon. As a
result of prompt fission, the muons are entrained on a fragment, and
then can be emitted due to internal conversion during the cascade
deexcitation of the fragments.

\begin{figure}
\hspace*{3cm}
\includegraphics[width=0.8\textwidth]{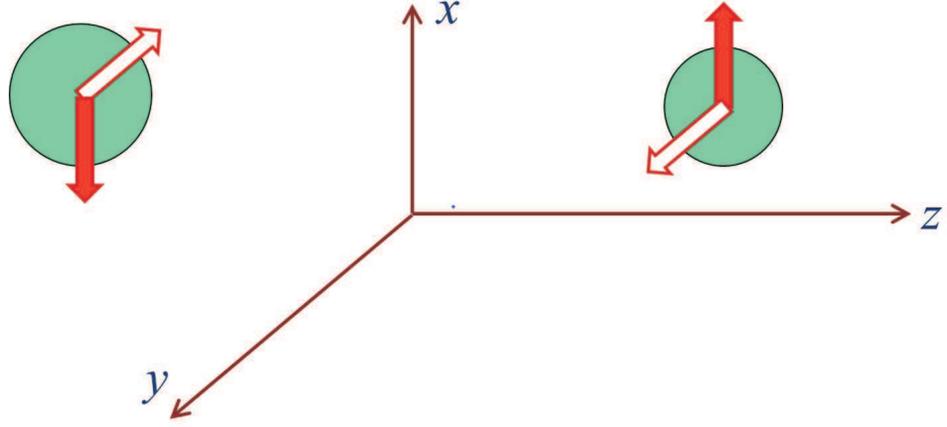}
\vspace*{2cm}
\caption{\footnotesize Scheme of the fragment spin alignment in the plane perpendicular to the fission axis. }
\label{f1}
\end{figure}

      Experimentally the angular distribution of the muons from prompt fission of $^{233}$U was undertaken in Refs. \cite{belov} in nuclear photoemulsions. Theoretical calculations were performed in \cite{kar84}. Calculated spectrum of the conversion muons is up to 1--2 MeV. Furthermore, experiment shows a significant focusing of the muons along the fission axis. This can be understood in terms of the alignment of the fragments in the plane perpendicular to the direction of fission \cite{skars} (Fig. \ref{f1}). Correspondingly, two ways of analysis can be put forward. One looks very natural: consider that the  spin of the fragment is directed somewhere in the $(x,y)$ plane, so that the spins of the muons are assumed $M = J$ in the internal system. After recalculation to the labsystem, one has to average over the directions of the spin in the azimuthal $(x,y)$ plane. Such would be a HV way. Another way, consecutively QM one, is that the quantization axis is chosen along the fission direction, in correspondence with the experimental conditions of observation. Then the spin projections of the muons are assumed $M = 0$.

      \section{Formulas}
      \label{corl}

     The actual angular distribution depends on the multipolarity of
the transition. Consider first $E1$ transitions. In the HV model,
the angular distribution of the emitted neutrons in the internal
coordinate system will be \be \Chi'(\theta', \phi') =
|Y_{11}(\theta',\phi')|^2 \sim \sin^2\theta' \,. \label{hv} \ee
After transformation to the angle $\alpha$ relative to the fission
axis $z$ in the c. m. system of the fragment and averaging over
the azimuthal angle $\phi$ of the spin in the $(x,y)$ plane, one
arrives at the angular distributions in this system, calculated
within the HV-theory: \be \Chi_{h.v.} (\alpha) =  \frac12
(1+\cos^2\alpha)  \,. \label{hvl} \ee Factor of $\frac12$ takes
into account the normalization of the full number of the emitted
particles. Contrary, within the framework of the QM approach one
obtains \be \Chi_{\text{QM}} (\alpha) =  |Y_{10} (\alpha,\phi)|^2
\sim \cos^2 \alpha \,. \label{QM} \ee

\begin{figure}
\hspace*{3em}
\includegraphics[width=0.39\textwidth]{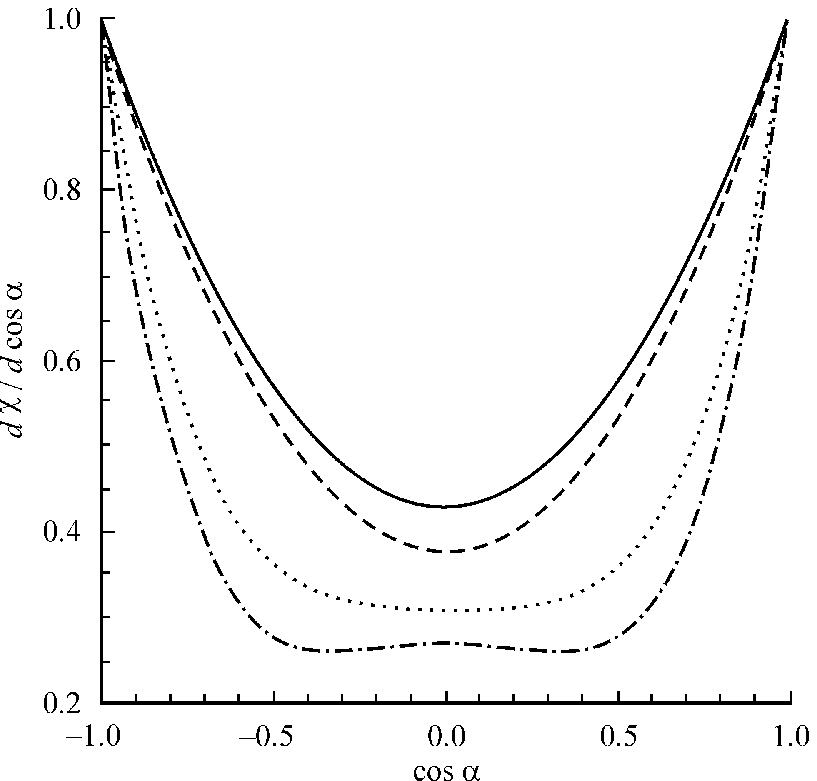}
\hspace*{2em}
\includegraphics[width=0.39\textwidth]{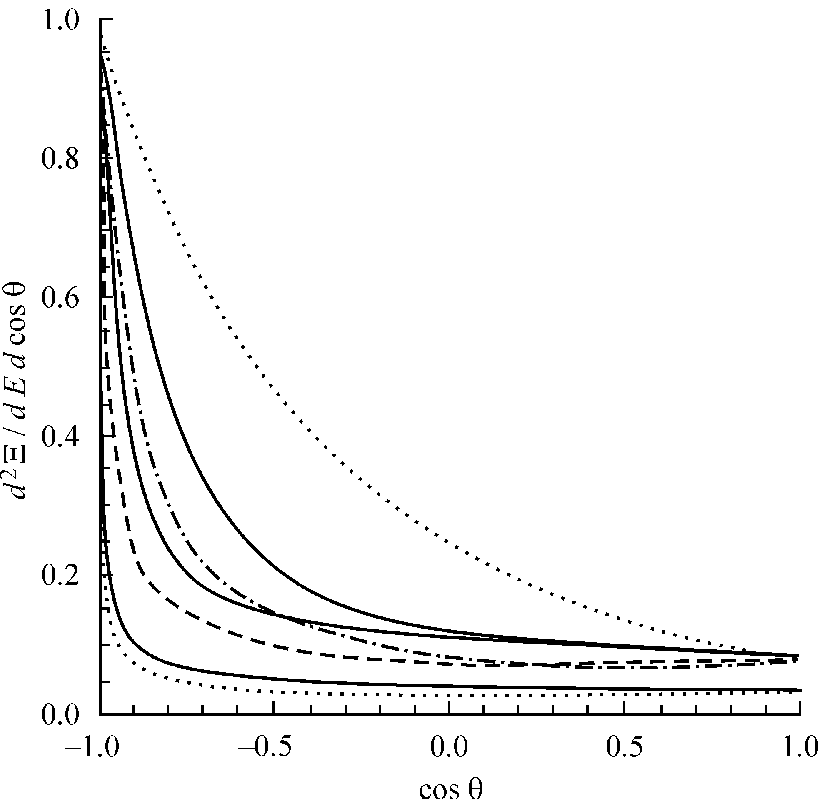}
\caption{\footnotesize Left: calculated angular distribution of the conversion muons in the center-of-mass system for the $E1$ transitions in the case of the initial spin of the nucleus $I$ = 7 and 4 (solid and dashed lines) and for the $E2$ transitions (dashed
and dash-dotted lines, respectively). In the case of isotropic emission,
$\frac{d\chi}{d\cos \alpha}= const$. Right: that of the conversion muons for light fragments with the atomic number $Z$ = 40 in the laboratory system, for the $E1$- and $E2$-transitions with the kinetic energy of the muons $E_\mu$ = 1, 0.25 and 0.1 MeV, respectively (in the order of approach to the origin). Upper dotted line corresponds to isotropic emission in the c.m. system.}
\label{resf}
\end{figure}

    In the case of the $E2$ transitions, one has the angular distribution in the c.m. system
$\Phi' (\theta', \phi') = |Y_{22} (\theta',\phi')|^2 \sim
\sin^4\theta$. Correspondingly, one obtains in the labsystem
expressions
\bea
\Phi_{h.v.} (\alpha) =  \frac{3\pi}4 (1+\frac23\cos^2\alpha +\cos^4 \alpha)  \label{hv2}  \\
 \Phi_{\text{QM}} (\alpha) =  |Y_{20} (\alpha,\phi)|^2 =
1-6 \cos^2 \alpha + 9\cos^4\alpha \,.    \label{QM2}
\eea
      Comparison of Eqs. (\ref{hvl}) and (\ref{QM}) shows that the HV polarization turns out to be twice less than the QM one (see next section in more detail). This conflict shows inconsistence of HV with quiantum mechanics. Furthermore, in the case of the $E2$ emission, the form itself of the angular dependence becomes different. Experiment can test which approach is true.  The difference speaks for itself.

      \section{Comparison to experiment}
      \label{contradic}

      For comparison with  experiment, it should be taken into account that the alignment of the fragments is not 100\%. Therefore, the general expression for the $(n, n)$ correlations instead of (\ref{hv}) will be written as
\be
\chi'(\theta', \phi') = 1+A_{nJ}\sin^2 \theta'\,,  \label{eq1}
\ee
and for the $ (n, f) $  correlation --- in the form
\be
\chi_{QM}(\alpha) = 1+A_{nf}\cos^2 \alpha \,  \label{QM3}
\ee
instead of (\ref{QM}). Experience shows that $A_{nf} \ll  1$ is a parameter within 10 \% \cite{cora}. Then in the labsystem, Eq. (\ref{hvl}) goes over
\be
\chi_{h.v.} (\alpha) = 1 + \frac12 A_{nJ}\cos^2\alpha  \,.
\label{hv3}
\ee
Comparing (\ref{hv3}) with (\ref{QM3}), in the case of the $E1$ transitions one arrives at the relation \cite{cora}
\be
 A_{nJ} \simeq 2 A_{nf}  \,.   \label{fac2}
\ee

      The results are presented in Fig. \ref{resf}, left for the both $E1$ and $E2$ transitions in the c.m. of a fragment. It is symmetric for the angles between 0 and $\pi$, and has a distinctive 0---$\frac \pi 2$ anisotropy. In the labsystem, the distribution concentrates along the velocity of the fragment, as shown in Fig. \ref{resf}, right.
\begin{figure}[!tbh]
\includegraphics[width=0.45\textwidth]{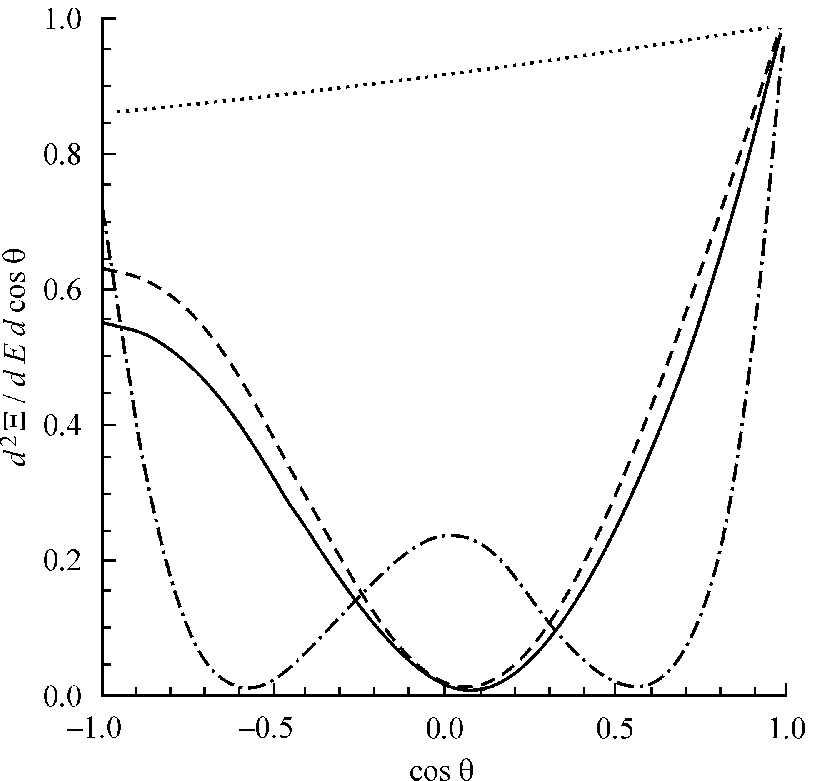}
\hspace*{3em}
\includegraphics[width=0.45\textwidth]{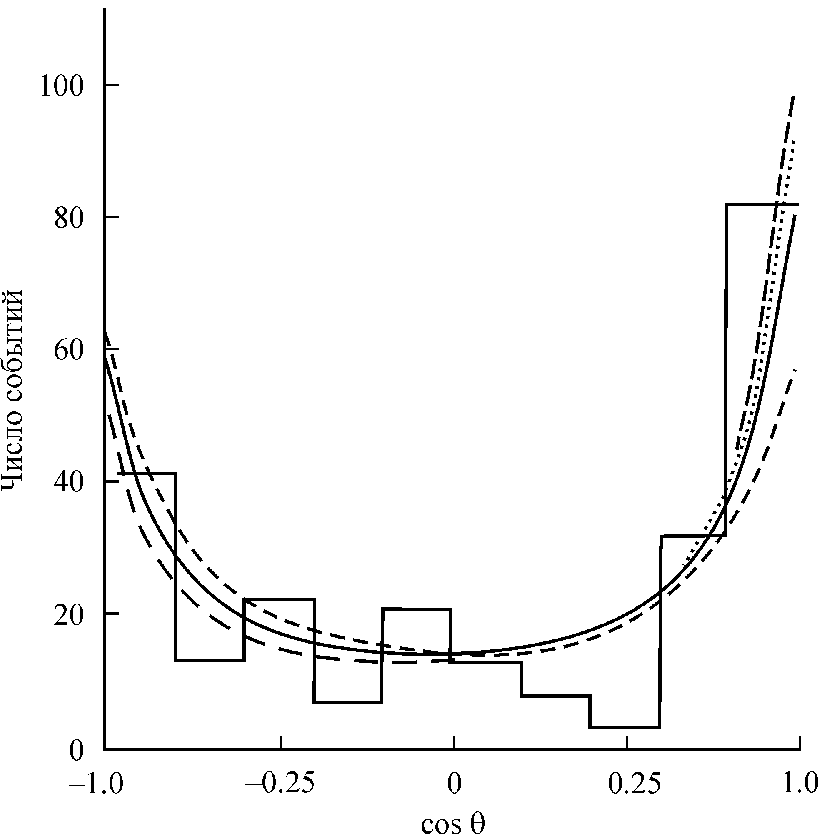}
\caption{\footnotesize Left: angular distribution of shake muons. Solid and dashed lines: the $E1$   transitions with the kinetic energy of the conversion muons $E_\mu$ = 1 and 0.5 MeV, respectively. Dotted and dash-dotted lines: the $E2$ and $E0$ transitions with $E_\mu$ = 1 MeV, respectively. Right: comparison of the calculated angular distribution with experiment \cite{belov}. The solid curve corresponds to a mixture of the multipoles 90\%$E1$ + 10\%$E2$,
dotted line --- 50\%$E1$ + 15\%$E2$ + 35\%$M1$. For comparison, short and long strokes show the angular distribution for pure $E1$ and $E2$ transitions, respectively. }
\label{expf}
\end{figure}

       Shake-off muons come from a sudden change of the potential due to snapping-back the remnants of the neck. They can be compared to the scission neutrons. As distinct from the latter, emission of the shake muons is sharply anisotropic (Fig. \ref{expf}, left).

      In Fig. \ref{expf}, right the theoretical angular distributions, averaged over the fragment distributions and the energies of the muons are presented in comparison with the experimental histogram from Ref. \cite{belov}. Two characteristics of the fission fragments can be inferred from this comparison. \\
1)There is a fraction of the $E2$ component at the level of 15\% in the radiation spectrum of the fission fragments. \\
2) There is also a share of the shake muons within 5\%. \\
Of course, these values are obtained within the QM approach,
without HV. Shake muons are similar to the scission neutrons.

\section{Conclusion}

$(\mu, f)$ as well as $(n,f)$ angular correlations appear as a
convenient example for experimental investigation of the difference between   the  HV theory and  QM approach. Alignment and polarization are two manifestations of the spin
state. They should be considered in different experiments.
Projection $M = J$ in the internal system of fragments does not mean that detection of another
projection of the spin --- on the perpendicular axis of fission  --- will
result in $M = 0$. Degree of polarization of such states can be
measured in $(n,n)$ angular correlations. The state $M=0$ is a
different vector of state. It should be measured
simultaneously with detection of the fission direction. This
result shows that if we knew the hidden spin projection, we could
calculate of course the angular distribution, also with respect to
any direction, {\it e.g.} fission direction. But in order to know,
one should first make a measurement of it. And  this is a
different experiment. Therefore, one should chose the quantization
axis in accordance with what is observed  experimentally, and
consider the corresponding magnetic quantum number of the spin.
Simultaneous consideration of another projection turns out to be
unobservable --- that is hidden variable, whose involving  into
calculation leads to a wrong result.

    At the same time, note that the two distributions, given by  Eqs. (\ref{hvl}) and (\ref{QM}), are presently under CORA experimental investigation \cite{cora} of the $(n,n)$ and $(n,n,f)$ correlations. Thus, the ratio of the parameters turns out to be a measurable quantity. The equality of this ratio to two is justified above only for the case of 100\% polarization or alignment. Moreover, it is different in the case of the $E2$ transitions. In general case, the angular distribution also depends on the initial and final spins of the fragments, as one can see from Fig. \ref{resf}. Therefore, the resulting form (\ref{hv3}) or (\ref{QM3}) instead of (\ref{hv2}) or (\ref{QM2}) of the emission spectra can be due to averaging over all the fragments with their spins and multipole mixtures of the transitions.

Furthermore, in the internal coordinate system, the fragments are 100\% polarized. Therefore, one could expect, that in the case of $(n, n)$ angular correlations, the right form should be just (\ref{hv}), not (\ref{eq1}), under supposition of mostly $E1$ transitions, though. And the choice of $A_{nJ}$ has nothing to do with $A_{nf}$ as taken from the $(n,f)$ experiment.
The CORA experiment, performed at IPHC Strasbourg, aims at elucidating neutron emission mechanisms in the fission. Experimental check of the relations obtained above, and Eq. (\ref{fac2}) first of all, looks very  significant for the progress to this end.

\bigskip

       The author would like to express his gratitude to I. Guseva for
detailed and fruitful discussions.


\end{document}